\documentclass[12pt]{article}
\usepackage{amssymb}
\usepackage{amsmath}
\usepackage{latexsym}
\usepackage{mathrsfs}
\usepackage[all]{xy}

\date{\nonumber}
\def \be {\begin{equation}}
\def \ee {\end{equation}}

\title{\Large\bf Integrable dispersionless KdV hierarchy with sources}
\author{\small Zhihua Yang\thanks{E-mail: yangzhihua@tsinghua.org.cn}\ ,\ Ting Xiao\ and Yunbo Zeng\\
{\small {Department of Mathematical Sciences, Tsinghua University,
Beijing 100084, P. R. China} } }

\begin{document}
\maketitle

\setlength{\baselineskip}{16pt}
\renewcommand{\theequation}{\arabic{section}.\arabic{equation}}
\renewcommand{\thesection}{\arabic{section}}
\parbox[b]{1.2truecm}{\quad}
\parbox[b]{13truecm}{}
\vspace{0.5truecm}

\parbox[b]{0.2truecm}{\quad}
\parbox[b]{13.8truecm}{\small
{\bf Abstract. } An integrable dispersionless KdV hierarchy with
sources (dKdVHWS) is derived. Lax pair equations and
bi-Hamiltonian formulation for dKdVHWS are formulated. Hodograph
solution for the dispersionless KdV equation with sources (dKdVWS)
is obtained via hodograph transformation. Furthermore, the
dispersionless Gelfand-Dickey hierarchy with sources (dGDHWS) is
presented.   \par PACS number: 02. 03. Ik}

 \section{\large Introduction}
 \setcounter{equation}{0}
 In recent years, research in the dispersionless hierarchies has become quite
 active (see, for example, \cite{JB00}-\cite{CT03I} and references therein).
 Dispersionless hierarchies arise as the quasiclassical limit of
 the original dispersionfull hierarchies \cite{TT95}. The operators
 in the Lax equations for dispersionfull hierarchy are replaced by phase
 functions for dispersionless hierarchy, commutators are replaced by Poisson brackets and the
 role of Lax pair equations by dispersionless Lax pair equations.
 The dispersionless hierarchies have Hamiltonian formulation \cite{JB00,JB96}
 and many other aspects \cite{CT03II,KZ02,AYN04}, and several methods of
 solutions of dispersionless hierarchies have been
 formulated \cite{YK88,YK89,XZ06I,XZ06II,CT03I}.

 \par
 The soliton equations with self-consistent sources (SESCS) are another
 type of integrable models and have important physical
 applications \cite{VM90I}-\cite{YZ99}. There are some ways to derive the
 SESCS, for example, the Mel'nikov way \cite{VM90I,VM88,VM92,VM89,VM90II} and the Leon
 approach \cite{JL88,LL90,JL90}. In recent years, SESCS were studied based
 on the constrained flows of soliton equations which are just the stationary
 equations of SESCSs\cite{MA92I,YZ94}. There are several methods for
 solving the SESCSs, for example, the inverse scattering method \cite{VM88,VM92,LZM01}, the matrix
 theory \cite{VM89}, the $\bar{\partial}$ method and gauge
 transformation \cite{DS95,SD96}, and the
 Darboux transformation \cite{YZ94,ZMS01,ZSX03,ZS05}.
 By treating the variable $x$ as the evolution parameter and $t$
 as the 'spatial' variable respectively, and introducing
 Jacobi-Ostrogradiski coordinates, the SESCS has a $t$-type
 Hamiltonian formulation \cite{MB95,YZ99}.
 \par
 This paper is devoted to the integrable dispersionless KdV
 hierarchy with sources (dKdVHWS). Considering the asymptotic
 expansion of the wave function, and taking the dispersionless limit
 of the KdV hierarchy with sources (KdVHWS), we can deduce the dKdVHWS.
 The Lax pair
 equations of the dKdVHWS can be deduced by the dispersionless
 limit of the Lax pair equations of the KdVHWS. Similar to
 the dKdV hierarchy, the dKdVHWS has bi-Hamiltonian formulation
 and can be solved via hodograph transformation. The
 Gelfand-Dickey hierarchy with sources (GDHWS) is the integrable
 generalization of the Gelfand-Dickey hierarchy, and the corresponding
 integrable dispersionless hierarchy, i.e. the dGDHWS, can be deduced.
 \par This paper is organized as follows: in Section 2 we review
 some definitions and results about the KdVHWS. In Section 3 we derive the
 dKdVHWS as well as its Lax pair equations by taking the dispersionless
 limit of the KdVHWS. In Section 4, we construct the
 bi-Hamiltonian formulation of dKdVHWS. Then we derive the
 hodograph solution for dispersionless KdV equation with sources (dKdVWS) in Section 5.
 In Section 6, we deduce the integrable
 dispersionless Gelfand-Dickey hierarchy with sources (dGDHWS).
 Some conclusion is made in Section 7.

 \section{\large The KdV hierarchy with sources}
 \setcounter{equation}{0} We first briefly review some definitions and
 results about the KdV hierarchy with sources in
 the framework of Sato theory. Given a
 pseudo-differential operator (PDO) of the form \cite{Di02}
 \be L=\partial^2+u \ee
 where $\partial=\frac{\partial}{\partial x}$, $u=u(x,t)$,
 $t=(t_3,t_5,\dots)$, and the wave function $\psi=\psi(x,t)$,
 consider the Lax pair
 \begin{subequations}
 \be L\psi=\lambda\psi,\ee \be \psi_{t_{2m-1}}=B_{2m-1}\psi,\ee
 \end{subequations}
 where $\lambda$ is a parameter, $B_{2m-1}=(L^{\frac{2m-1}{2}})_+$
 , $m=1,2,3,\dots$, and $(A)_+$ here standing for the
 differential part of $A$. The compatibility condition of
 $(2.2\mbox{a})$ and $(2.2\mbox{b})$ gives rise to \be \frac{\partial L}{\partial
 t_{2m-1}}=[B_{2m-1},L], \ee which is the well-known KdV hierarchy\cite{Di02}.
 As was shown in \cite{Di02}, the KdV hierarchy could be written as bi-Hamiltonian
 systems \be u_{t_{2m-1}}=B_0\frac{\delta \mathscr{H}_{2m+1}}{\delta u}=B_1
 \frac{\delta
 \mathscr{H}_{2m-1}}{\delta u},\ m=1,2,\dots, \ee where
 $\mathscr{H}_{2m-1}=\int h_{2m-1}dx$ is a functional of $u$,
 $B_0=\partial$, $B_1=\frac{1}{4}\partial^3+u\partial+\frac{1}{2}u_x$
 are Hamiltonian operators,
 and $\frac{\delta \mathscr{H}_{2m-1}}{\delta u}$
 is the Euler-Lagrange derivative of the
 Hamiltonian $\mathscr{H}_{2m-1}$ defined as
 \be\frac{\delta \mathscr{H}_{2m-1}}{\delta u}
 =\frac{\delta}{\delta u}\int h_{2m-1}dx=\sum_{k=0}^{\infty}
 (-\partial)^k\frac{\partial h_{2m-1}}{\partial u^{(k)}}.\ee

 \par Let us consider \cite{MA92I} \be \widetilde{B}_{2m-1}=B_{2m-1}+\sum_{k=1}^n
 \psi_k\partial^{-1}\phi_k,\ \ m=1,2,3,\dots,\ee where
 $\psi_k=\psi_k(x,t)$ and $\phi_k=\phi_k(x,t)$ satisfying \be L\psi_k=\lambda_k\psi_k,\
 L^*\phi_k=\lambda_k\phi_k,\ k=1,\dots,n,\ee here $L^*=(-\partial)^2+u=L$ is the
 adjoint operator of $L$, and $\lambda_k$ is a constant, $k=1,\dots,n$.
 Then the KdV hierarchy with sources (KdVHWS)
 \cite{MA92I,LZM01,ZMS01,VM90II} can be defined as
 \begin{subequations}\be \frac{\partial L}{\partial
 t_{2m-1}}=[\widetilde{B}_{2m-1},L],\ee
 \be L\psi_k=\lambda_k\psi_k,\ee\be L^*\phi_k=\lambda_k\phi_k
 \ee\end{subequations}
 with the Lax pair given by
 \begin{subequations}
 \be L\psi=\lambda\psi,\ee \be \psi_{t_{2m-1}}=\widetilde{B}_{2m-1}\psi,\ee
 \end{subequations}
 namely, under $(2.8\mbox{b})$ and $(2.8\mbox{c})$, the compatibility condition
 of $(2.9\mbox{a})$ and $(2.9\mbox{b})$
 gives rise to $(2.8\mbox{a})$.

 \section{\large Dispersionless limit}
 \setcounter{equation}{0}
 Following the procedure introduced in \cite{TT95,JB96,XZ06I,XZ06II}, we could
 derive the dispersionless hierarchy by taking the dispersionless
 limit of the initial system. Taking $T=\epsilon t,\ X=\epsilon x,$ and
 thinking of $u(x,t)=u(\frac{X}{\epsilon},\frac{T}{\epsilon})=
 U(X,T)+O(\epsilon)$ as $\epsilon\rightarrow 0$, $L$ in
 $(2.1)$ changes into \be L_{\epsilon}=\epsilon^2\partial_X^2+
 u(\frac{X}{\epsilon},\frac{T}{\epsilon})=
 \epsilon^2\partial_X^2+U(X,T)+O(\epsilon),\ee
 where $\partial_X=\frac{\partial}{\partial X}$.
 It can be proved \cite{JB96} that \be \mathcal{L}=
 \sigma^{\epsilon}(L_{\epsilon})=p^2+U\ee  satisfies
 \be \mathcal{L}
 _{T_{2m-1}}=\{\mathcal{B}_{2m-1},\mathcal{L}\},\ee
 where $\sigma^{\epsilon}$
 denotes the principal symbol \cite{TT95}, the bracket $\{,\}$ is the Poison bracket
 defined in $2D$ 'phase space' $(p,X)$ as \be \{A,B\}=\frac{\partial A}{\partial
 p}\frac{\partial B}{\partial X}
 -\frac{\partial A}{\partial X}\frac{\partial B}{\partial p},\ee
 and $\mathcal{B}_{2m-1}=(\mathcal{L}^{\frac{2m-1}{2}})_+$ now refers to
 nonnegative powers of $p$.
 We define $(3.3)$ as the dispersionless KdV (dKdV) hierarchy
 \cite{JB96}, and the first few equations are expressed as
 \begin{subequations}
 \be U_{T_1}=U_X,\ee \be U_{T_3}=\frac{3}{2}UU_X,\ee \be
 U_{T_5}=\frac{15}{8}U^2U_X,\ee \be
 U_{T_7}=\frac{35}{16}U^3U_X,\dots\ee
 \end{subequations}
 \par As was shown in \cite{JB96}, equation $(3.5\mbox{b})$ has
 tri-Hamiltonian formulation as
 \be U_{T_3}=\mathcal{D}_1\frac{\delta H_5}{\delta U}
 =\mathcal{D}_2\frac{\delta H_3}{\delta U}
 =\frac{3}{4}\mathcal{D}_3\frac{\delta H_1}{\delta U},\ee
 where
 $$\mathcal{D}_1=2\partial_X,\ \mathcal{D}_2=U\partial_X+\partial_XU,\
 \mathcal{D}_3=U^2\partial_X+\partial_XU^2,$$
 $$H_1=\int UdX,\ H_3=\frac{1}{4}\int U^2dX,\ H_5=\frac{1}{8}\int U^3dX.$$
 For the general case, define\be H_{2m-1}=\frac{2}{2m-1}Tr\mathcal{L}^{\frac{2m-1}{2}},\ee
 where $TrA=\int ResAdX$, and $ResA$ is the residue of the general
 Laurent polynomial of the form
 $A=\sum_{i=-\infty}^{+\infty}a_i(X)p^i$, i.e. the coefficient of
 the $p^{-1}$ term, then the Hamiltonians $(3.7)$ are in
 involution with respect to any of the three Poisson brackets
 \be \{H_{2m-1},H_{2l-1}\}_i=\int dX\frac{\delta H_{2m-1}}{\delta U}\mathcal{D}_i
 \frac{\delta H_{2l-1}}{\delta U}=0,\ \ i=1,2,3, \ee
 and dKdV hierarchy $(3.3)$ have
 tri-Hamiltonian formulation
 \be U_{T_{2m-1}}=\mathcal{D}_1\frac{\delta H_{2m+1}}{\delta U}=\mathcal{D}_2\frac{\delta H_{2m-1}}{\delta
 U}=\frac{(2m-1)(2m-3)}{(2m-2)^2}\mathcal{D}_3\frac{\delta H_{2m-3}}{\delta
 U},\ m=2,3,\dots.\ee
 \par It was also shown in \cite{JB96} that the
 solution of $(3.5\mbox{b})$ can be described through the implicit form
 \be U=f(X+\frac{3}{2}UT_3),\ee
 where $f$ is an arbitrary function.
 \par
 In what follows, we derive the dispersionless KdV hierarchy
 with sources (dKdVHWS).
 \par By taking $T=\epsilon t,\ X=\epsilon x,$ $(2.8)$ change into
 \begin{subequations}\be
 \epsilon L_{\epsilon T_{2m-1}}=[B_{\epsilon(2m-1)}+\sum_{k=1}^n\psi_k(\frac{X}{\epsilon},\frac{T}{\epsilon})
 (\epsilon\partial_X)^{-1}\phi_k(\frac{X}{\epsilon},\frac{T}{\epsilon}),L_{\epsilon}],\ee\be
 L_{\epsilon}\psi_k(\frac{X}{\epsilon},\frac{T}{\epsilon})=\lambda_k\psi_k(\frac{X}{\epsilon},\frac{T}{\epsilon}),\ee\be
 L^*_{\epsilon}\phi_k(\frac{X}{\epsilon},\frac{T}{\epsilon})=\lambda_k\phi_k(\frac{X}{\epsilon},\frac{T}{\epsilon}),
 \ee\end{subequations}
 where $B_{\epsilon(2m-1)}=(L_{\epsilon}^{\frac{2m-1}{2}})_+$.
 \par Similar to the dispersionless KP case in \cite{TT95,XZ06I,XZ06II},
 we consider the following WKB asymptotic expansion of $\psi_k(\frac{X}{\epsilon},\frac{T}{\epsilon})$ and
 $\phi_k(\frac{X}{\epsilon},\frac{T}{\epsilon})$, $k=1,2,\dots,n$,
 \begin{subequations}
 \be \psi_k(\frac{X}{\epsilon},\frac{T}{\epsilon})\sim exp\{\frac{S(X,T,\lambda=\lambda_k)}
 {\epsilon}+\beta_{k1}+O(\epsilon)\},\ \epsilon\rightarrow 0,\ee
 \be \phi_k(\frac{X}{\epsilon},\frac{T}{\epsilon})\sim exp\{-\frac{S(X,T,\lambda=\lambda_k)}
 {\epsilon}+\beta_{k2}+O(\epsilon)\},\ \epsilon\rightarrow 0,\ee
 \end{subequations}
 it can be calculated that
 \begin{eqnarray}&&\psi_k(\frac{X}{\epsilon},\frac{T}{\epsilon})
 (\epsilon\partial_X)^{-1}\phi_k(\frac{X}{\epsilon},\frac{T}{\epsilon})\nonumber\\
 &=& e^{\beta_{k1}+\beta_{k2}}[(\epsilon\partial_X)^{-1}+(\frac{\partial}{\partial X}S(X,T,\lambda=\lambda_k)+O(\epsilon))
 (\epsilon\partial_X)^{-2}\nonumber\\
 &&+((\frac{\partial}{\partial X}S(X,T,\lambda=\lambda_k))^2+O(\epsilon))
 (\epsilon\partial_X)^{-3}+\dots], \nonumber
 \end{eqnarray}
 as $\epsilon\rightarrow0$. Therefore, we have \be \sigma^{\epsilon}
 \Big{(}\psi_k(\frac{X}{\epsilon},\frac{T}{\epsilon})(\epsilon\partial_X)^{-1}
 \phi_k(\frac{X}{\epsilon},\frac{T}{\epsilon})\Big{)}=\frac{v_k}{p-p_k},\ee
 where \be v_k=e^{\beta_{k1}+\beta_{k2}},\ p_k=\frac{\partial}{\partial X}S(X,T,\lambda=\lambda_k).\ee
 Taking the principal symbol of both sides of $(3.11\mbox{a})$, we have
 \be \mathcal{L}
 _{T_{2m-1}}=\{\mathcal{B}_{2m-1}+\sum_{k=1}^n\frac{v_k}{p-p_k},\mathcal{L}\}
 =\{\mathcal{B}_{2m-1},\mathcal{L}\}+\{\sum_{k=1}^n\frac{v_k}{p-p_k},\mathcal{L}\},\ee
 and the dispersionless limit of $(3.11\mbox{b})$ and $(3.11\mbox{c})$
 lead to \be p_k^2+U=\lambda_k,\ \frac{\partial}{\partial X}(p_kv_k)=0.\ee
 \par Under $(3.16)$, it can be found that
 \be\{\frac{v_k}{p-p_k},\mathcal{L}\}=-2v_{k,X},\ee
 therefore, the dispersionless limit of $(3.11)$, i.e. dKdVHWS,
 reads \begin{subequations}\be
 U_{T_{2m-1}}=\{\mathcal{B}_{2m-1},\mathcal{L}\}-2\sum_{k=1}^nv_{k,X},
 \ m=1,2,3,\dots,\ee
 \be p_k^2+U=\lambda_k,\ee
 \be \frac{\partial}{\partial X}(p_kv_k)=0.\ee
 \end{subequations}
 Integrating $(3.18\mbox{c})$ and taking $\lambda_k^m$ as the integral
 constants, then we can eliminate $v_k$ in $(3.18\mbox{a})$ and
 rewrite dKdVHWS in another form
 \be U_{T_{2m-1}}=\{\mathcal{B}_{2m-1},\mathcal{L}\}-2\sum_{k=1}^n
 \Big{(}\frac{\lambda_k^m}{\sqrt{\lambda_k-U}}\Big{)}_X.\ee
 \par If we take the dispersionless limit of $(2.9)$, we will
 obtain the Lax pair equations of dKdVHWS $(3.18)$ as
 \begin{subequations}\be p^2+U=\lambda,\ee\be
 p_{T_{2m-1}}=\Big{(}\mathcal{B}_{2m-1}+\sum_{k=1}^n\frac{v_k}{p-p_k}\Big{)}_X
 ,\ee\end{subequations}
 namely, under $(3.18\mbox{b})$ and $(3.18\mbox{c})$, the compatibility condition
 of $(3.20\mbox{a})$ and $(3.20\mbox{b})$ gives rise to $(3.18\mbox{a})$. We
 can eliminate $v_k$ and $p_k$ in $(3.20\mbox{b})$ and rewrite
 $(3.20)$ in another form
 \begin{subequations}\be p^2+U=\lambda,\ee\be
 p_{T_{2m-1}}=\Big{(}\mathcal{B}_{2m-1}+\sum_{k=1}^n
 \frac{\lambda_k^m}{p\sqrt{\lambda_k-U}-(\lambda_k-U)}\Big{)}_X,\ee\end{subequations}
 which are the Lax pair equations of $(3.19)$.
 \par We give two examples in the following, the first one is the dispersionless KdV equation with sources
 (dKdVWS)
 \be U_{T_3}=\frac{3}{2}UU_X-2\sum_{k=1}^n
 \Big{(}\frac{\lambda_k^2}{\sqrt{\lambda_k-U}}\Big{)}_X,\ee
 with Lax pair equations given by
 \begin{subequations}\be p^2+U=\lambda,\ee\be
 p_{T_3}=\Big{(}p^3+\frac{3}{2}Up+\sum_{k=1}^n
 \frac{\lambda_k^2}{p\sqrt{\lambda_k-U}-(\lambda_k-U)}\Big{)}_X.\ee\end{subequations}
 And the second example is the dispersionless KdV(5) equation with sources
 (dKdV(5)WS)
 \be U_{T_5}=\frac{15}{8}U^2U_X-2\sum_{k=1}^n
 \Big{(}\frac{\lambda_k^3}{\sqrt{\lambda_k-U}}\Big{)}_X,\ee
 with Lax pair equations given by
 \begin{subequations}\be p^2+U=\lambda,\ee\be
 p_{T_5}=\Big{(}p^5+\frac{5}{2}Up^3+\frac{15}{8}U^2p+\sum_{k=1}^n
 \frac{\lambda_k^3}{p\sqrt{\lambda_k-U}-(\lambda_k-U)}\Big{)}_X.\ee\end{subequations}

 \section{\large Hamiltonian formulation of dKdVHWS}
 \setcounter{equation}{0} It is well known that the KdV hierarchy
 has bi-Hamiltonian formulation \cite{Di02}, the dKdV hierarchy has
 tri-Hamiltonian formulation \cite{JB96}, or even further, quasi-Hamiltonian
 formulation \cite{JB00}, and the KdVHWS has bi-Hamiltonian formulation
 \cite{MB95}. Motivated by the Hamiltonian formulation of the
 dKdV case \cite{JB96}, we would construct the
 bi-Hamiltonian formulation of the dKdVHWS $(3.19)$.
 \par Let us firstly consider dKdVWS $(3.22)$. Set
 \be A_k=2\lambda_k^2\int\sqrt{\lambda_k-U}\ dX,B_k=2\lambda_k\int\sqrt{\lambda_k-U}\ dX,\ee
 then by direct computation we have
 $$\mathcal{D}_1\frac{\delta A_k}{\delta U}=2\partial_X\Big{(}-\frac{\lambda_k^2}{\sqrt{\lambda_k-U}}\Big{)}
 =-2\Big{(}\frac{\lambda_k^2}{\sqrt{\lambda_k-U}}\Big{)}_X,$$
 $$\mathcal{D}_2\frac{\delta B_k}{\delta U}=(U\partial_X+\partial_XU)\Big{(}-\frac{\lambda_k}{\sqrt{\lambda_k-U}}\Big{)}
 =-\frac{\lambda_k^2U_X}{(\lambda_k-U)^{3/2}}=-2\Big{(}\frac{\lambda_k^2}{\sqrt{\lambda_k-U}}\Big{)}_X.$$
 Therefore, if we denote
 \begin{subequations}\be\widetilde{H}_3=H_3+\sum_{k=1}^nB_k=\int
 dX\big{(}\frac{1}{4}U^2+2\sum_{k=1}^n\lambda_k\sqrt{\lambda_k-U}\big{)},\ee
 \be\widetilde{H}_5=H_5+\sum_{k=1}^nA_k=\int
 dX\big{(}\frac{1}{8}U^3+2\sum_{k=1}^n\lambda_k^2\sqrt{\lambda_k-U}\big{)},\ee\end{subequations}
 then equation $(3.22)$ can be written in two Hamiltonian forms
 \be U_{T_3}=\mathcal{D}_1\frac{\delta \widetilde{H}_5}{\delta U}=\mathcal{D}_2\frac{\delta \widetilde{H}_3}{\delta
 U}.\ee
 \par For the dKdVHWS $(3.19)$,
 denote \be \widetilde{H}_{2m-1}=H_{2m-1}+2\sum_{k=1}^n\lambda_k^{m-1}\int\sqrt{\lambda_k-U}\ dX,\ee
 we can directly prove (see $Appendix$) that the Hamiltonians $\widetilde{H}_{2m-1},\ m=1,2,\dots$
 satisfy
 \be \{\widetilde{H}_{2m-1},\widetilde{H}_{2l-1}\}_i=\int dX\frac{\delta \widetilde{H}_{2m-1}}{\delta U}\mathcal{D}_i
 \frac{\delta \widetilde{H}_{2l-1}}{\delta U}=0,\ \ i=1,2, \ee
 therefore, the dKdVHWS $(3.19)$ have bi-Hamiltonian formulation
 \be U_{T_{2m-1}}=\mathcal{D}_1\frac{\delta \widetilde{H}_{2m+1}}{\delta U}=
 \mathcal{D}_2\frac{\delta \widetilde{H}_{2m-1}}{\delta
 U},\ m=1,2,\dots.\ee

 \section{\large Hodograph solution for dKdVWS}
 \setcounter{equation}{0} In this section, using the hodograph
 transformation\cite{YK88,XZ06I,XZ06II}, we will derive the hodograph
 solution for the the dKdVWS $(3.22)$.
 Following \cite{YK88} and let $U_{T_3}=B(U)U_X$, we can find
 from $(3.22)$ that \be B(U)=\frac{3}{2}U-\sum_{k=1}^n\frac{\lambda_k^2}{(\lambda_k-U)^{3/2}}.\ee
 Making the hodograph transformation with the change of variables
 $(X,T_3)\rightarrow(U,T_3)$ and let $X=X(U,T_3)$, we have
 \be0=\frac{dX}{dT_3}=\frac{\partial X}{\partial U}\frac{\partial U}{\partial T_3}
 +\frac{\partial X}{\partial T_3}=\frac{\partial X}{\partial U}BU_X+\frac{\partial X}{\partial
 T_3},\ee
 which implies that \be \frac{\partial X}{\partial T_3}=-B
 =-\frac{3}{2}U+\sum_{k=1}^n\frac{\lambda_k^2}{(\lambda_k-U)^{3/2}}.\ee
 It can be integrated as \be X+\Big{(}\frac{3}{2}U-\sum_{k=1}^n\frac{\lambda_k^2}{(\lambda_k-U)^{3/2}}\Big{)}T_3=F(U),\ee
 where $F(U)$ is an arbitrary function of $U$, $(5.4)$ gives an
 implicit solution of $(3.22)$.
  \par When $F=0$ and $\lambda_1=\lambda_2=\dots=\lambda_n=0$,
 dKdVWS $(3.22)$ degenerates to dKdV equation $(3.5\mbox{b})$, and $(5.4)$
 degenerates to the rational solution of $(3.5\mbox{b})$, $U=-\frac{2X}{3T}$.
 \par When $F\neq0$ and is convertible, then solution of dKdVWS $(3.22)$
 can be written through the implicit form
 \be
 U=F^{-1}(X+\Big{(}\frac{3}{2}U-\sum_{k=1}^n
 \frac{\lambda_k^2}{(\lambda_k-U)^{3/2}}\Big{)}T_3),\ee
 which is similar to $(3.10)$.

 \section{\large The dispersionless  Gelfand-Dickey hierarchy with sources}
 \setcounter{equation}{0}
 The well-known Gelfand-Dickey hierarchy with sources (GDHWS) \cite{MA92I}
 is defined as
 \begin{subequations}\be
 \frac{\partial L}{\partial
 t_m}=[\widetilde{B}_m,L]=[B_m+\sum_{k=1}^n
 \psi_k\partial^{-1}\phi_k,L],\ee\be
 L\psi_k=\lambda_k\psi_k,\ee\be L^*\phi_k=\lambda_k\phi_k
 ,\ee\end{subequations}
 where \be L=\partial^N+u_{N-2}\partial^{N-2}+\dots+u_1\partial+u_0,\ee
 $u=(u_{N-2},\dots,u_0)^T$, $u_i=u_i(x,t)$, $i=0,1,\dots,N-2$, $t=(t_2,t_3,\dots)$,
 $B_m=[(L^{\frac{1}{N}})^m]_+$,
 $L^*=(-\partial)^N+(-\partial)^{N-2}u_{N-2}+\dots+(-\partial)u_1+u_0$
 is the adjoint operator of $L$, $\lambda_k$ is a constant, $k=1,\dots,n$,
 $\psi_k=\psi_k(x,t)$ and $\phi_k=\phi_k(x,t)$,
 and the Lax pair is given by
 \begin{subequations}
 \be L\psi=\lambda\psi,\ee \be \psi_{t_m}=\widetilde{B}_m\psi,\ee
 \end{subequations}
 namely, under $(6.1\mbox{b})$ and $(6.1\mbox{c})$, the compatibility condition of
 $(6.3\mbox{a})$ and $(6.3\mbox{b})$ gives rise to $(6.1\mbox{a})$.
 \par Following the procedure given above, we can derive the
 dispersionless Gelfand-Dickey hierarchy with sources (dGDHWS).
 Taking $T=\epsilon t$, $X=\epsilon x$, and let
 $u_k(\frac{X}{\epsilon},\frac{T}{\epsilon})=U_k(X,T)+O(\epsilon)$
 as $\epsilon\rightarrow0$, then $(6.1)$ change into
 \begin{subequations}\be
 \epsilon L_{\epsilon T_m}=[B_{\epsilon m}+\sum_{k=1}^n
 \psi_k(\frac{X}{\epsilon},\frac{T}{\epsilon})
 (\epsilon\partial_X)^{-1}\phi_k(\frac{X}{\epsilon},\frac{T}{\epsilon}),L_{\epsilon}],\ee\be
 L_{\epsilon}\psi_k(\frac{X}{\epsilon},\frac{T}{\epsilon})=
 \lambda_k\psi_k(\frac{X}{\epsilon},\frac{T}{\epsilon}),\ee\be
 L_{\epsilon}^*\phi_k(\frac{X}{\epsilon},\frac{T}{\epsilon})=
 \lambda_k\phi_k(\frac{X}{\epsilon},\frac{T}{\epsilon})
 ,\ee\end{subequations}
 where \begin{eqnarray}L_{\epsilon}&=&(\epsilon\partial_X)^N+
 u_{N-2}(\frac{X}{\epsilon},\frac{T}{\epsilon})(\epsilon\partial_X)^{N-2}+\dots+
 u_1(\frac{X}{\epsilon},\frac{T}{\epsilon})\epsilon\partial_X+
 u_0(\frac{X}{\epsilon},\frac{T}{\epsilon})\nonumber\\
 &=&(\epsilon\partial_X)^N+
 (U_{N-2}(X,T)+O(\epsilon))(\epsilon\partial_X)^{N-2}+\dots\nonumber\\
 &&\quad+(U_1(X,T)+O(\epsilon))\epsilon\partial_X
 +U_{0}(X,T)+O(\epsilon),\end{eqnarray}
 and $B_{\epsilon m}=[(L_{\epsilon}^{\frac{1}{N}})^m]_+.$
 Consider the following WKB asymptotic expansion of
 $\psi_k(\frac{X}{\epsilon},\frac{T}{\epsilon})$ and
 $\phi_k(\frac{X}{\epsilon},\frac{T}{\epsilon})$, $k=1,\dots,n$,
 \begin{subequations}
 \be \psi_k(\frac{X}{\epsilon},\frac{T}{\epsilon})\sim exp\{\frac{S(X,T,\lambda=\lambda_k)}
 {\epsilon}+\beta_{k1}+O(\epsilon)\},\ \epsilon\rightarrow 0,\ee
 \be \phi_k(\frac{X}{\epsilon},\frac{T}{\epsilon})\sim exp\{-\frac{S(X,T,\lambda=\lambda_k)}
 {\epsilon}+\beta_{k2}+O(\epsilon)\},\ \epsilon\rightarrow 0,\ee
 \end{subequations}
 then the principal symbol \cite{TT95} of $(6.4\mbox{a})$ arises
  \be \frac{\partial\mathcal{L}}
 {\partial T_m}=\{\mathcal{B}_m+\sum_{k=1}^{n}\frac{v_k}{p-p_k},\mathcal{L}\}\ee
 where \be\mathcal{L}=
 \sigma^{\epsilon}(L_{\epsilon})=p^N+U_{N-2}p^{N-2}+\dots+U_1p+U_0,\ee $\mathcal{B}_m=
 \sigma^{\epsilon}(B_{\epsilon m}),$ and $v_k=e^{\beta_{k1}+\beta_{k2}},\ p_k=
 \frac{\partial}{\partial X}S(X,T,\lambda=\lambda_k)$ are
 obtained by\be \sigma^{\epsilon}
 \Big{(}\psi_k(\frac{X}{\epsilon},\frac{T}{\epsilon})(\epsilon\partial_X)^{-1}
 \phi_k(\frac{X}{\epsilon},\frac{T}{\epsilon})\Big{)}=\frac{v_k}{p-p_k}.\ee
 The dispersionless limit of $(6.4\mbox{b})$ and $(6.4\mbox{c})$ give
 rise to
 \begin{subequations}\be p_k^N+U_{N-2}p_k^{N-2}+\dots+U_1p_k+U_0=\lambda_k,\ee
 \be \frac{\partial}{\partial X}\Big{(}v_k\cdot\frac{\partial\mathcal{L}}{\partial
 p}\Big{|}_{p=p_k}\Big{)}=0,\ee
 \end{subequations}
 $(6.7)$ together with $(6.10\mbox{a})$ and $(6.10\mbox{b})$ give rise to the
 dispersionless Gelfand-Dickey hierarchy with sources (dGDHWS)
 \begin{subequations}\be
 \frac{\partial\mathcal{L}}
 {\partial
 T_m}=\{\mathcal{B}_m+\sum_{k=1}^{n}\frac{v_k}{p-p_k},\mathcal{L}\},\ee\be
 p_k^N+U_{N-2}p_k^{N-2}+\dots+U_1p_k+U_0=\lambda_k,\ee\be
 \frac{\partial}{\partial X}\Big{(}v_k\cdot\frac{\partial\mathcal{L}}{\partial
 p}\Big{|}_{p=p_k}\Big{)}=0,\ee\end{subequations}
 whose Lax pair equations are given by
 \begin{subequations}\be
 p^N+U_{N-2}p^{N-2}+\dots+U_1p+U_0=\lambda,\ee\be
 p_{T_m}=\left(\mathcal{B}_m+\sum_{k=1}^n\frac{v_k}{p-p_k}\right)_X,\ee\end{subequations}
 namely, under $(6.11\mbox{b})$ and $(6.11\mbox{c})$, the compatibility condition of
 $(6.12\mbox{a})$ and $(6.12\mbox{b})$ gives rise to $(6.11\mbox{a})$.
 \par Under $(6.11\mbox{b})$ and $(6.11\mbox{c})$, it can be found
 by a tedious computation that
 \be\{\frac{v_k}{p-p_k},\mathcal{L}\}=a_{N-2}p^{N-2}+a_{N-3}p^{N-3}+\dots+a_0,\ee
 where $a_{N-2}=-Nv_{k,X}$, $a_{N-3}=-N(v_kp_k)_X$, and
 for $i=4,\dots,N$, \be
 a_{N-i}=-v_k\sum_{j=1}^{i-3}jp_k^{j-1}U_{N+1-i+j,X}-\sum_{l=2}^{i-2}
 (N-l)(v_kp_k^{i-2-l})_XU_{N-l}-N(p_k^{i-2}v_k)_X.\ee
 When $N=2$, we have $\mathcal{L}=p^2+U$, and $(6.13)$ is the
 same as $(3.17)$.
 \par Similar to the dKdVHWS, the dGDHWS possess bi-Hamiltonian
 formation and their solutions can be obtained via hodograph
 transformation.

 \section{\large Conclusion}
 \setcounter{equation}{0}
 We derive dKdVHWS by taking the dispersionless limit of KdVHWS,
 meanwhile, Lax pair equations of the dKdVHWS can be obtained by taking the
 dispersionless limit of the corresponding dispersionfull equations. We have
 constructed the bi-Hamiltonian formulation of the dKdVHWS, and have
 obtained the implicit solutions of the dKdVWS
 via the hodograph transformation. For the generalization case, we
 have deduced the dGDHWS which also possess bi-Hamiltonian formulation
 and can be solved via the hodograph transformation.

 \section*{ Acknowledgment}\hskip\parindent
 This work was supported by the Chinese Basic Research Project
 "Nonlinear Science". \hskip\parindent

 \vspace{3cm}
 \section*{Appendix}
 \setcounter{equation}{0}
 Here we give the proof of involution relation of the Hamiltonians
 $\widetilde{H}_{2m-1}$ $(4.4)$.
 \par Since $$\{\widetilde{H}_{2m-1},\widetilde{H}_{2l-1}\}_1
 =\{\widetilde{H}_{2m-1},\widetilde{H}_{2l-3}\}_2
 =-\{\widetilde{H}_{2l-3},\widetilde{H}_{2m-1}\}_2
 =-\{\widetilde{H}_{2l-3},\widetilde{H}_{2m+1}\}_1$$
 $$=\{\widetilde{H}_{2m+1},\widetilde{H}_{2l-3}\}_1
 =\dots=\{\widetilde{H}_{2m+2l-3},\widetilde{H}_1\}_1,$$
 it suffices to prove that for any $m\geq1$,
 $\{\widetilde{H}_{2m-1},\widetilde{H}_1\}_1=0$. We can directly
 calculate that
 \begin{eqnarray}
 &&\{\widetilde{H}_{2m-1},\widetilde{H}_1\}_1\nonumber\\
 &=&\int\frac{\delta \widetilde{H}_{2m-1}}{\delta U}\mathcal{D}_1
 \frac{\delta \widetilde{H}_1}{\delta U}dX\nonumber\\
 &=&\int\Big{(}\frac{\delta H_{2m-1}}{\delta U}
 -\sum_{k=1}^n\lambda_k^{m-1}\frac{1}{\sqrt{\lambda_k-U}}\Big{)}
 2\partial_X\Big{(}1-\sum_{k=1}^n\frac{1}{\sqrt{\lambda_l-U}}\Big{)}dX\nonumber\\
 &=&-\int\Big{(}\frac{\delta H_{2m-1}}{\delta U}
 -\sum_{k=1}^n\lambda_k^{m-1}\frac{1}{\sqrt{\lambda_k-U}}\Big{)}
 \sum_{l=1}^n\frac{U_X}{\sqrt{\lambda_l-U}(\lambda_l-U)}dX.\nonumber
 \end{eqnarray}
 Since all $\frac{\delta H_{2m-1}}{\delta U}$, $m=1,2,\dots$ are of the form
 $cU^s$, where $c$ are constants, and $s\in \mathcal{N}$, it suffices to prove
 $$\int\frac{U^sU_X}{\sqrt{\lambda_l-U}(\lambda_l-U)}dX=0,\quad l=1,\dots,n,\eqno(I)$$
 $$\int\frac{U_X}{\sqrt{\lambda_k-U}\sqrt{\lambda_l-U}(\lambda_l-U)}dX=0,\quad
 k,l=1,\dots,n.\eqno(II)$$
 \par For $(I)$, let $F_l=\sqrt{\lambda_l-U}$, then $U=\lambda_l-F_l^2$,
 $U_X=-2F_lF_{l,X}$ and
 $$\int\frac{U^sU_X}{\sqrt{\lambda_l-U}(\lambda_l-U)}dX
 =-2\int\frac{(\lambda_l-F_l^2)^s}{F_l^2}dF_l,$$
 which are all zero for the reason that
 $\frac{(\lambda_l-F_l^2)^s}{F_l^2}$
 are rational polynomials of $F_l^2$ and so
 $\frac{(\lambda_l-F_l^2)^s}{F_l^2}dF_l$ are total differentials.
 \par For $(II)$, when $k=l$, then $(II)$ obviously holds;
 when $k\neq l$, let $G_l=\lambda_l-U$, then $U_X=-G_{l,X}$, and
 \begin{eqnarray}
 &&\int\frac{U_X}{\sqrt{\lambda_k-U}\sqrt{\lambda_l-U}(\lambda_l-U)}dX\nonumber\\
 &=&\int\frac{-G_{l,X}}{\sqrt{\lambda_k-\lambda_l+G_l}\sqrt{G_l}G_l}dX\nonumber\\
 &=&\int\frac{-G_{l,X}}{G_l^2\sqrt{\frac{\lambda_k-\lambda_l}{G_l}+1}}dX\nonumber\\
 &=&\frac{1}{\lambda_k-\lambda_l}\int\frac{1}{\sqrt{\frac{\lambda_k-\lambda_l}{G_l}+1}}
 d(\frac{\lambda_k-\lambda_l}{G_l}),\nonumber
 \end{eqnarray}
 set $\frac{\lambda_k-\lambda_l}{G_l}=tan^2H_{kl}$, then
 $1+\frac{\lambda_k-\lambda_l}{G_l}=sec^2H_{kl}$,
 $d(\frac{\lambda_k-\lambda_l}{G_l})=2tanH_{kl}sec^2H_{kl}dH_{kl}$, and
 $$\frac{1}{\sqrt{\frac{\lambda_k-\lambda_l}{G_l}+1}}d(\frac{\lambda_k-\lambda_l}{G_l})
 =\frac{2tanH_{kl}sec^2H_{kl}dH_{kl}}{secH_{kl}}=2d(secH_{kl}),$$
 which are total differentials, and this proves $(II)$.


\vspace{2cm}

\begin{thebibliography}{s99}
\bibitem{JB00}
Brunelli J C 2000 {\em Braz. J. Phys.} \textbf{30} 455
\bibitem{TT95}
Takasaki T and Takebe T 1995 {\em Rev. Math. Phys.} \textbf{7} 743
\bibitem{JB96}
Brunelli J C 1996 {\em Rev. Math. Phys.} \textbf{8} 1041
\bibitem{CT03II}
Chen Y T and Tu M H 2003 {\em Lett. Math. Phys.} \textbf{63} 125
\bibitem{KZ02}
Zheltukhin K 2002 {\em Phys. Lett. A} \textbf{297} 402
\bibitem{AYN04}
Alagesan T Chung Y and Nakkeeran K 2004 {\em Chaos, Solitons and
Fractals} \textbf{21} 63
\bibitem{YK88}
Kodama Y 1988 {\em Phys. Lett. A} \textbf{129} 223
\bibitem{YK89}
Kodama Y 1989 {\em Phys. Lett. A} \textbf{135} 171
\bibitem{XZ06I}
Xiao T and Zeng Y B 2006 {\em Phys. Lett. A} \textbf{349} 128
\bibitem{XZ06II}
Xiao T and Zeng Y B 2006 {\em J. Nlin. Math. Phys.} at press
\bibitem{CT03I}
Chen Y T and Tu M H 2003 {\em J. Phys. A: Math. Gen.} \textbf{36}
9875
\bibitem{VM90I}
Mel'nikov V K 1990 {\em J. Math. Phys.} \textbf{31} 1106
\bibitem{JL88}
Leon J1988 {\em J. Math. Phys.} \textbf{29} 2012
\bibitem{LL90}
Leon J and Latifi A 1990 {\em J. Phys. A} \textbf{23} 1385
\bibitem{JL90}
Leon J 1990 {\em Phys. Lett. A} \textbf{144} 444
\bibitem{MA92I}
Antonowicz M 1992 {\em Phys. Lett. A} \textbf{165} 47
\bibitem{YZ94}
Zeng Y B 1994 {\em Phys. D} \textbf{73} 171
\bibitem{VM88}
Mel'nikov V K 1988 {\em Phys. Lett. A} \textbf{133} 493
\bibitem{VM92}
Mel'niklov V K 1992 {\em Phys. Lett. A} \textbf{8} 133
\bibitem{LZM01}
Lin R L, Zeng Y B and Ma W X 2001 {\em Phys. A} \textbf{291} 287
\bibitem{VM89}
Mel'nikov V K 1989 {\em Comm. Math. Phys.} \textbf{120} 451
\bibitem{DS95}
Doktorov E V and Shchesnovich V S 1995 {\em Phys. Lett. A}
\textbf{207} 153
\bibitem{SD96}
Shchesnovich V S and Doktorov E V 1996 {\em Phys. Lett. A}
\textbf{213} 23
\bibitem{ZMS01}
Zeng Y B, Ma W X and Shao Y S 2001 {\em J. Math. Phys.}
\textbf{42} 2113
\bibitem{ZSX03}
Zeng Y B, Shao Y J and Xue W M 2003 {\em J. Phys. A: Math. Gen.}
\textbf{36} 5035
\bibitem{ZS05}
Zeng Y B, Shao Y J 2005 {\em J. Phys. A: Math. Gen.} \textbf{38}
2441
\bibitem{MB95}
Blaszak M 1995 {\em J. Math. Phys.} \textbf{36} 4826
\bibitem{YZ99}
Zeng Y B 1999 {\em Phys. A} \textbf{262} 405
\bibitem{Di02}
Dickey L A 2002 {\em Soliton equations and Hamiltonian systems
(2nd ed.)}(World Scientific: Singapore)
\bibitem{VM90II}
Mel'nikov V K 1990 {\em Nonlinear evolution equations and
dynamical systems} (Berlin: Springer)
\end {thebibliography}
\end{document}